\documentclass[twocolumn,aps,prl,superscriptaddress]{revtex4-2}

\usepackage{amsmath}
\usepackage{amssymb}
\usepackage{amsfonts}
\usepackage{graphicx}
\usepackage{MnSymbol}

\begin{document}

\title{Rubber adhesion and friction: role of surface energy and contamination films}

\author{A. Tiwari}
\address{MultiscaleConsulting, Wolfshovener Str 2, 52428 J\"ulich, Germany}
\author{T. Tolpekina}
\address{Apollo Tyres Global R\&D B.V., P.O. Box 3795, 7500 DT | Colosseum 2, 7521 PT Enschede, The Netherlands}
\author{Hans van Benthem}
\address{Apollo Tyres Global R\&D B.V., P.O. Box 3795, 7500 DT | Colosseum 2, 7521 PT Enschede, The Netherlands}
\author{M.K. Gunnewiek}
\address{Apollo Tyres Global R\&D B.V., P.O. Box 3795, 7500 DT | Colosseum 2, 7521 PT Enschede, The Netherlands}
\author{B.N.J. Persson}
\address{MultiscaleConsulting, Wolfshovener Str 2, 52428 J\"ulich, Germany}
\address{PGI-1, FZ J\"ulich, Germany}

\begin{abstract}
We study the influence of the surface energy and contamination films on rubber adhesion and sliding friction.
We find that there is a transfer of molecules from the rubber to the substrate which
reduces the work of adhesion and makes the rubber friction insensitive to the substrate surface energy.
We show that there is no simple relation between adhesion and friction: 
adhesion is due to (vertical) detachment processes
at the edge of the contact regions (opening crack propagation), 
while friction in many cases is determined 
mainly by (tangential) stick-slip instabilities of nanosized regions, within the whole
sliding contact. Thus while the pull-off force in fluids may be strongly reduced (due to a reduction of the
work of adhesion), the sliding friction may be only slightly affected
as the area of real contact may be dry, and the frictional shear stress 
in the contact area nearly unaffected by the fluid.
\end{abstract}

\maketitle

\pagestyle{empty}


{\bf 1 Introduction}

The friction and adhesion between rubber materials and a counter surface have many
practical applications, e.g., for tires, conveyor belts, rubber seals and pressure sensitive
adhesives. The adhesion and friction of rubber is a very complex topic 
because of the temperature and frequency dependency, and the non-linearity, of the
stress-strain relation. In addition, all solids have surface roughness, 
usually extending from the linear size of the object down to
atomic distances, which strongly influence the contact 
mechanics\cite{Ref1,Ref2,Ref3,Ref4,SSR,Klupp1,Gert11,Cq}.  

In this paper we will study the influence of the substrate surface energy on adhesion and friction. 
We use surfaces with the same topography but different surface energy, namely smooth and 
sandblasted glass surfaces without and with hydrophobic coating (monolayers of grafted molecules).
We will also study how the transfer of rubber to the track may influence the friction for another 
rubber compound on the surface contaminated by the first compound.

This paper is organized as follows. In Sec. 2 we present a qualitative discussion about rubber friction.
In Sec. 3 we describe the rubber compounds used in the study. In Sec. 4 
we show the surface roughness power spectra of the sandblasted glass surfaces,
and of a smooth and a rough rubber surface. Sec. 5 describe the procedure used to prepare hydrophobic
glass surfaces. In Sec. 6 we present adhesion results between clean and silanized glass
balls and rubber, which illustrate the influence 
on the adhesion by contamination films, formed by transfer of molecules 
from the rubber to the glass surface.
In Sec. 7 we present Fourier-transform infrared spectroscopy data
which shows that the contamination film is a wax, which is added 
to the rubber to protect against ozone.
Similarly, during sliding of rubber blocks, contamination films (derived from the rubber) 
form on the countersurface, even for the silanized glass surfaces.
This is discussed in Sec. 8 where we present
experimental results for the velocity dependent friction coefficient 
for rubber blocks sliding on smooth and sandblasted glass surfaces, both ``clean'' and silanized.
We will also study how the transfer of molecules from a rubber compound A to a concrete surface influence the friction for another 
rubber compound B on the surface contaminated by the compound A.
Sec. 9 present a discussion, and Sec. 10 the summary and conclusion.

\begin{figure}[tbp]
\includegraphics[width=0.47\textwidth,angle=0]{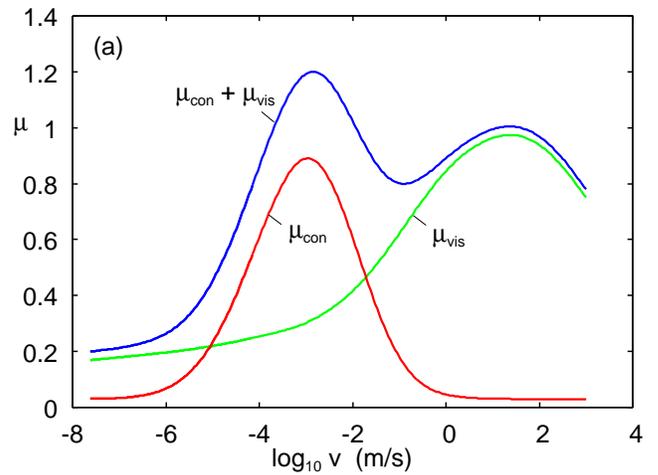}
\caption{
The contribution to the friction coefficient from viscoelastic deformations $\mu_{\rm vis}$, and from the
area of real contact $\mu_{\rm con}$, as function of sliding 
speed in a typical case at room temperature. 
}
\label{Schematic.1logv.2mu.eps}
\end{figure}

\begin{figure}[!ht]
\includegraphics[width=0.95\columnwidth]{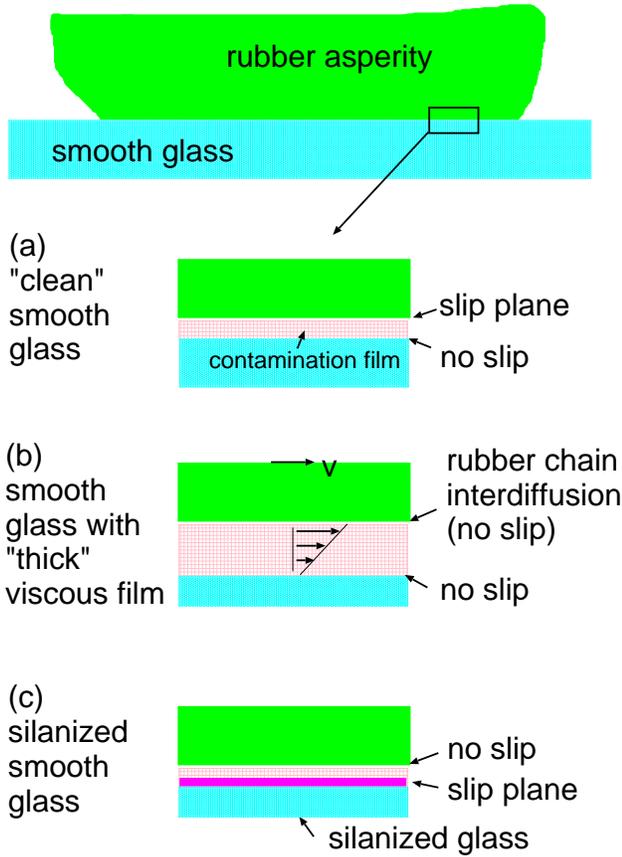}
\caption{\label{DiffCases.eps}
The contribution to the friction from the area of real contact depends on 
surface energies, contamination films and slip (or shear) planes. In (a) a thin strongly bound contamination
film occur on the substrate surface.
In (b) a thicker contamination film occur, e.g. due to transfer of molecules from the 
rubber to the glass surface. In (c) the glass surface is silanized which
may result in slip between the contamination film and the silanized glass surface (pink).
}
\end{figure}

\vspace{0.3cm}
{\bf 2 Rubber friction: qualitative discussion}

Fig. \ref{Schematic.1logv.2mu.eps} shows schematically
the contribution to the friction coefficient $\mu = \mu_{\rm vis}+\mu_{\rm con}$
from viscoelastic deformations $\mu_{\rm vis}$, and from the
area of real contact $\mu_{\rm con}$, as function of sliding 
speed in a typical case of rubber sliding on a rough rigid countersurface at room temperature\cite{Res}. 

The contribution from the area of real contact may involve shearing a thin contamination film, or
processes where polymer segments (or nanosized rubber patches) undergoes cyclic stick-slip events\cite{Schall,Leonov,theory3}, or where
hard filler particles at the rubber surface scratches the substrate (plowing friction).
The viscoelastic contribution to the friction result from the time-dependent deformations of the rubber by
the substrate asperities, and occurs on many length scales\cite{Persson9}. 
In what follows we will focus mainly on the contribution to the friction from the area of real contact.

We can write $\mu_{\rm con} = \tau_{\rm f} A$, where the area of real contact $A$ depends on the contact pressure,
temperature and the sliding speed. The frictional shear stress $\tau_{\rm f}$ can originate from different
processes which we now discuss. 

Fig. \ref{DiffCases.eps} shows schematically the contribution to the friction from the area of real contact,
which depends on surface energies, contamination films and the location of slip (or shear) planes.
In (a) a thin, strongly bound, contamination
film (usually water and polar molecules with a hydrocarbon part) occur on the surface.
In (b) a thicker contamination film occur, which could result from transfer of molecules from the rubber 
to the glass surface. In (c) the glass surface is silanized, and hence covered by an inert grafted monolayer 
(pink region), which may result in slip between the contamination film and the silanized surface.

If the contamination film is very thin (some nanometers), and the molecules bound strongly enough to the substrate
surface (here glass), slip will occur between the rubber and the contamination film (see Fig. \ref{DiffCases.eps}(a)).
In this case the contribution from the area $A$ of real contact to the friction may involve
interfacial processes where polymer segments, or nanosized patches of the rubber, attach, stretch, snap-off and 
re-attach to the substrate\cite{Schall,Leonov,theory3}. During the rapid snap-off the elastic energy in the 
stretched polymer chains are converted into heat. Theoretical\cite{Schall,Leonov,theory3} and experimental\cite{add1,add2} 
studied have shown that this
result in a frictional shear stress which is well approximated by a Gaussian-like function of 
the logarithm of the sliding speed $v$, with a width of order several decades in velocity:
$$\tau_{\rm f} \approx \tau_0 \ {\rm exp} \left (-c\left [ {\rm log} {v\over v^*} \right ]^2 \right )\eqno(1)$$
where $v^*$ is a reference velocity which depends on the temperature but which for $T\approx 20^\circ {\rm C}$
typically is of order $1 \ {\rm mm/s}$.

If the contamination film is thicker (see Fig. \ref{DiffCases.eps}(b))
it may behave as a (usually shear thinning) viscous fluid,
and the shear stress acting on the rubber surface will be determined by the viscosity (and shear rate) of the
fluid:
$$\tau_{\rm f} \approx {\eta_0 \dot \gamma \over 1+(\eta_0/B) \dot \gamma^n}\eqno(2)$$ 
where $B$ is a shear-thinning constant 
and $\dot \gamma = v/d$ the shear rate ($d$ is the film thickness). 
For polymer fluids the exponent $n$ is typically in the range $0.7-1$ \cite{Yamada,IonPRL},
where the larger $n$ correspond to longer chain molecules.

For a silanized glass surface (see Fig. \ref{DiffCases.eps}(c)) some of the slip may be localized
to the interface between the hydrophobic coating and the contamination film\cite{recent},
but the experimental results presented below indicate that the main slip again occur either within the
contamination film or at the interface between the rubber and the contamination film. We will
show below that the contamination film may be derived mainly from the rubber rather than from other origins.

If the substrate is perfectly smooth there is no viscoelastic contribution $\mu_{\rm vis}$ to the friction.
This is the case even if the rubber surface has roughness. Several of the experiments reported on below use very smooth
glass surfaces, and in these cases $\mu \approx \mu_{\rm con}$.

\vspace{0.3cm}
{\bf 3 Rubber compounds}

We have used 3 different rubber compounds denoted A, B and C.
B is a summer tread compound filled with carbon black. 
A is a summer tread compound filled with silica and containing a traction resin. 
C is a winter tread compound,
filled with silica and containing a traction resin (same as in A in the same quantity).

Table \ref{table_glass_trans} gives a summary of the glass transition temperature $T_{\rm g}$ 
and the maximum of ${\rm tan}\delta$ for all the compounds as obtained 
from Dynamic Machine Analysis (DMA) measurements
of the viscoelastic modulus (see Ref. \cite{Tol}). 
The values in the parenthesis are from measurements performed one year earlier
on nominally identical rubber compounds from a different batch.

\begin{table}[hbt]
\caption{Summary of the glass transition temperatures of the 
A, B and C compounds. The glass transition temperature is defined as the
maximum of $\rm{tan} \delta$
as a function of temperature for the frequency $\omega_{0} = 0.01 \ \rm{s}^{-1}$.
The values in the parenthesis are from measurements performed one year earlier
on nominally identical rubber compounds from a different batch\cite{Tol}.}
\label{table_glass_trans}
\begin{center}
      \begin{tabular}{@{}|l||c|c|c|c|c|@{}}
\hline
compound & $T_{\rm g}$  & maximum of $\rm{tan} \delta$\\
\hline
\hline
A & $-28.2^\circ {\rm C}$ ($-30.4^\circ {\rm C}$) & 0.61 (0.53)\\
\hline
B & $-39.4^\circ {\rm C}$  ($-40.5^\circ {\rm C}$) & 0.57 (0.52)\\
\hline
C & $-46.2^\circ {\rm C}$ ($-47.2^\circ {\rm C}$) & 0.46 (0.41)\\
\hline
\end{tabular}
   \end{center}
\end{table}

\begin{figure}[!ht]
\includegraphics[width=0.9\columnwidth]{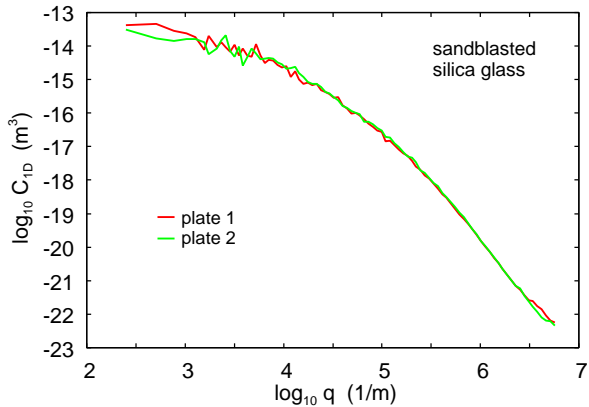}
\caption{\label{1logq.2logC1D.sandblastedGLASS.eps}
The 1D surface roughness power spectrum of the sandblasted glass plate 1 (red) and glass plate 2 (green).
The rms roughness of both plates is $\approx 15 \ {\rm \mu m}$.
}
\end{figure}

\begin{figure}[!ht]
\includegraphics[width=0.9\columnwidth]{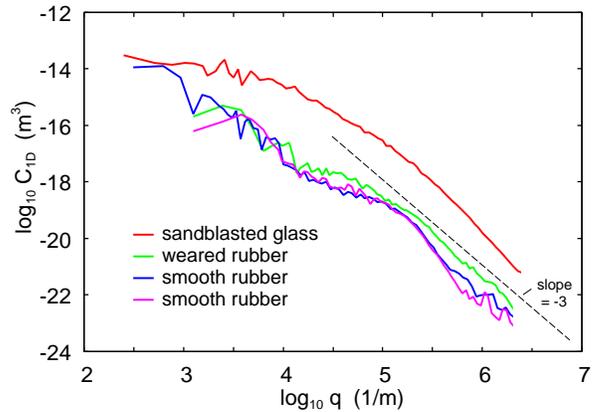}
\caption{\label{1logq.2logC.all.new.eps}
The 1D surface roughness power spectrum of the sandblasted glass surface (red) and of the
smooth (before tests) rubber surface (blue and pink), and of the rubber surface after the sliding friction measurements on the
sandblasted surface (green). 
}
\end{figure}

\vspace{0.3cm}
{\bf 4 Surface roughness power spectrum}

The surface roughness is of crucial importance for the sliding friction\cite{Cq,R1}.
The most important quantity is the surface roughness power spectrum. The two-dimensional (2D) surface roughness power spectrum $C(q)$,
which enters in the Persson contact mechanics theory, 
can be obtained from the height profile $z=h(x,y)$ measured over a square surface unit. However, for surfaces with
roughness with isotropic statistical properties, the 2D power spectrum 
can be calculated from the one-dimensional (1D) power spectrum obtained
from a line-scan $z=h(x)$ (see \cite{Nayak,Carbone}). 

The smooth glass surface can be considered as perfectly smooth with vanishing surface roughness power spectrum.
We prepared several sandblasted glass surfaces. The glass plates ($16 \ {\rm cm} \times 36  \ {\rm cm}$ and $0.5 \ {\rm cm}$ thick)
where sandblasted ``by hand'' for 5 minutes each. This may result in some variation in the surface topography but 
the surface roughness power spectrum of the plates are nearly identical (see Fig. \ref{1logq.2logC1D.sandblastedGLASS.eps}).
The power spectra shown in Fig. \ref{1logq.2logC1D.sandblastedGLASS.eps} was calculated from $25 \ {\rm mm}$ long line scans, 
and averaged over 5 different line scans obtained at different locations on the glass surfaces.

We also measured the surface topography of a rubber surface before and after sliding it on the sandblasted glass
surface. Sliding on the sandblasted surface resulted in some wear and increased surface roughness.
Fig. \ref{1logq.2logC.all.new.eps} shows the power spectra of the
smooth (before tests) rubber surface (blue and pink), and of the rubber surface after the sliding friction measurements on the
sandblasted surface (green). 
The figure also shows the surface roughness power spectrum of the sandblasted glass surface (red), as obtained from the average
over the two power spectra shown in Fig. \ref{1logq.2logC1D.sandblastedGLASS.eps}. 
The root-mean-square (rms) roughness of the sandblasted glass surface is $\approx 15 \ {\rm \mu m}$.

\begin{figure}[tbp]
\includegraphics[width=0.30\textwidth,angle=0]{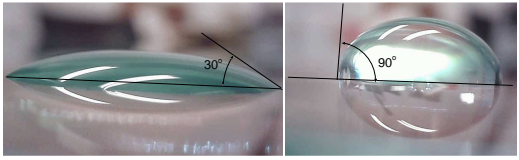}
\caption{
Water contact angle of glass surface cleaned with acetone and isopropanol (left),
and silanized glass surface i.e., a glass surface covered by hydrophobic grafted monolayer (right).
}
\label{hydrophob.hydrpphil.ps}
\end{figure}

\vskip 0.3cm
{\bf 5 Preparation of silanized glass surfaces} 

We have prepared silanized (hydrophobic) glass surfaces using Rain-X.
Rain-X is most commonly used on glass automobile surfaces, and result in a hydrophobic coating which causes water to bead.
Rain-X's primary active ingredient are polysiloxanes, 
the primary one being hydroxy-terminated polydimethylsiloxane. The polysiloxanes have functional groups that bind 
to the hydroxyl group of the glass surface. In addition Rain-X contains acetone and water, but the exact ratio is a trade secret. 

The glass surfaces was first cleaned with acetone and isopropanol, and then 
ultrasonic cleaned in distilled water. After drying the surface with a paper towel
we immediately applied the Rain-X. Next the glass surface
was wiped with a paper towel for 5 minutes using a firm, circular and overlapping motion.
Finally, the coated surface was left for 24 hours, washed with distilled water and dried. 
In addition, in a few studies the silanized glass surfaces were ultrasonic cleaned in distilled water to make sure just a monolayer
of chemically attached polydimethylsiloxane molecules occur on the glass surface.

Fig. \ref{hydrophob.hydrpphil.ps}
shows the water contact angle on a glass surface cleaned with acetone and isopropanol (left),
and a glass surface covered by hydrophobic monolayer (right) using Rain-X.

\begin{figure}
\includegraphics[width=0.45\textwidth,angle=0.0]{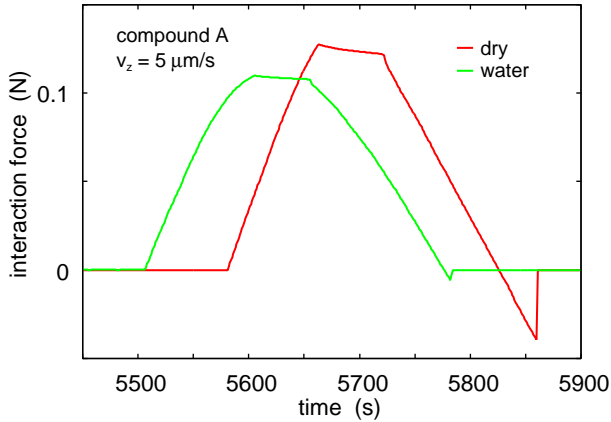}
\caption{\label{1time.2force.compound.A.eps}
The interaction force between a glass ball and the rubber compound A as a function of time.
Results are shown in dry condition (red) and when immersed in water (green).
The rubber surface was cleaned with hot water and the glass ball with acetone.
The glass ball moved up and down with the speed $5 \ {\rm \mu m/s}$.
}
\end{figure}

\begin{figure}
\includegraphics[width=0.45\textwidth,angle=0.0]{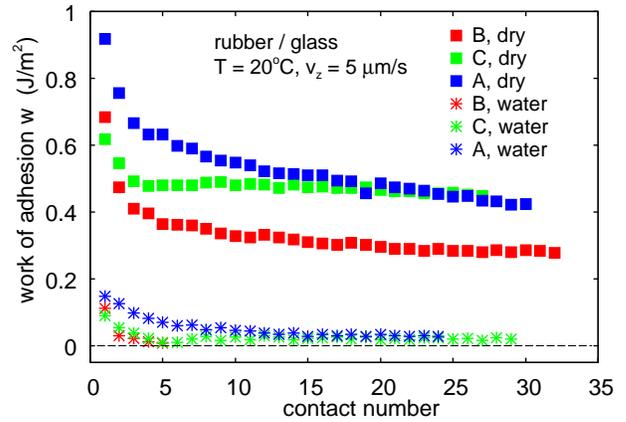}
\caption{\label{1number.2w.APOLLO.eps}
The work of adhesion between a glass ball and the rubber compound B, C and A for dry contact and in water.
The rubber surfaces are smooth and cleaned with hot water just before start of the experiments. The
glass ball was cleaned with acetone.
The pull-off speed  $v_z = 5 \ {\rm \mu m/s}$ and the temperature $T=20^\circ {\rm C}$.
}
\end{figure}

\begin{figure}
\includegraphics[width=0.45\textwidth,angle=0.0]{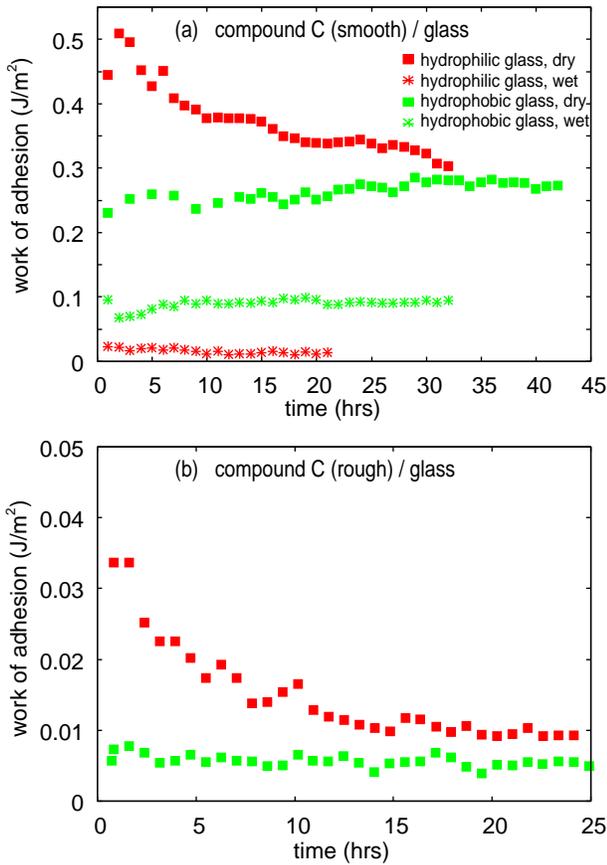}
\caption{\label{AdhesionCamoothrough.eps}
The work of adhesion between a glass ball and the rubber compound C for dry contact (squares) and in water (stars).
In (a) the rubber surface is smooth and cleaned with hot water just before start of the experiment. In (b) the rubber surface was
roughened by sandpaper just before start of the experiment. The red symbols are for a glass ball cleaned with acetone and isopropanol.
The green symbols are for a silanized glass ball prepared using Rain-X.
The pull-off speed  $v_z = 5 \ {\rm \mu m/s}$ and the temperature $T=20^\circ {\rm C}$.
}
\end{figure}

\vspace{0.3cm}
{\bf 6 Role of contamination on adhesion}

We have performed adhesion experiments 
where a glass ball with a few cm diameter 
is moved repeatedly up and down making first 
repulsive contact with the rubber substrate (typical repulsive force
$0.1 \ {\rm N}$) and then pulled-off\cite{Tol}.
The glass ball was cleaned with acetone just before start of the
experiment, and moved up and down with the vertical speed $5.0 \ {\rm \mu m/s}$.

The rubber surface is either smooth (but with some roughness due to the roughness of the mold)
or roughened by sandpaper P180. The smooth rubber surfaces were cleaned with hot water just before
the start of the experiments. This removed a gray film from the rubber surface (probably wax, see Sec. 7),
and made the rubber surfaces appear shining black,
but after a few hours the rubber surface appeared gray again.
This shows that mobile molecules 
diffuse to the rubber surface from inside the rubber, which probably 
is the origin of the time dependency of the adhesion observed below.

Fig. \ref{1time.2force.compound.A.eps}
shows the interaction force between a glass ball and the rubber compound A as a function of time.
Results are shown in dry condition (red) and when immersed in water (green).

Fig. \ref{1number.2w.APOLLO.eps}
shows the work of adhesion between an (originally) cleaned glass ball and the rubber compound 
B, C and A for dry contact and in water.
Note the continuous decrease in the work of adhesion with the number of contacts (and hence increasing time) which we attribute to
diffusion of mobile molecules to the rubber surface and to
transfer of molecules from the rubber to the glass surface. This is observed even in water. Note also that adhesion in
water is much smaller than in the dry state.

We have performed a set of experiments using rubber from a new (nominally identical) 
batch produced one year after the rubber used in Fig. \ref{1number.2w.APOLLO.eps}.
Fig. \ref{AdhesionCamoothrough.eps}
shows the work of adhesion between a glass ball and the rubber compound C for dry contact (squares) and in water (stars).
In (a) the rubber surface is smooth and cleaned with hot water just before start of the experiment. In (b) the rubber surface was
roughened by sandpaper just before start of the experiment. The red symbols are for a glass ball cleaned with acetone and isopropanol.
The green symbols are for a silanized glass ball prepared using Rain-X (see Sec. 5).

Note that for the roughened rubber surface the (macroscopic) adhesion nearly vanish. This is due to the reduction in the area
of real contact, and (more importantly) due to the elastic deformation energy stored at the interface which is ``given back''
during pull-off, and help to break the adhesive bonds. However, 
even when there is no macroscopic pull-off force, at short enough length scale adhesion is still strong,
and result in an increase in the contact area\cite{BoAmontons}. 

Fig. \ref{AdhesionCamoothrough.eps}(a) shows that for the smooth rubber surface 
the work of adhesion in the dry state is initially much larger for the clean glass ball than for the 
silanized glass ball, which we mainly attribute to a larger surface energy of the clean glass ball then the silanized ball.
However, the rubber surface is not perfectly smooth and the surface roughness will affect (reduce) the work of adhesion
and this effect may differ for the two cases.

For the (initially) clean glass ball the work of adhesion decreases with increasing number of contacts which we
attribute to a decrease in the surface energy of the glass ball due to transfer of molecules from the rubber to the glass ball.
For the silanized glass ball the work of adhesion is nearly constant,
both in the dry state and in water. This could indicate that no molecules are transferred from the
rubber to the silanized glass ball. However, with increasing number of contacts 
the work of adhesion for the (originally clean) glass ball approach the work of adhesion for the
silanized glass ball, so that even if molecules are transferred to the silanized glass ball they 
may result in a negligible change in the ball surface energy. Since the transferred molecules most likely are wax molecules
(see Sec. 7) this is not unexpected due to the inert nature of wax.

In water there is a huge difference between the
clean glass ball and the silanized glass ball. For the clean glass ball the adhesion nearly vanish which result from the hydrophilic
nature of the glass surface (water molecules bind strongly to the glass surface)
which, in the case of complete wetting, favor a thin (nanometer or less) 
water film between the glass and the rubber in the rubber-glass contact region, resulting in nearly vanishing adhesion.
The silanized glass surface is hydrophobic and the water does not like (from an energetic point of view) to stay between the rubber and the
glass surface, i.e., a dewetting transition occurs, resulting in strong adhesion even in water.

\begin{figure}
\includegraphics[width=0.45\textwidth,angle=0.0]{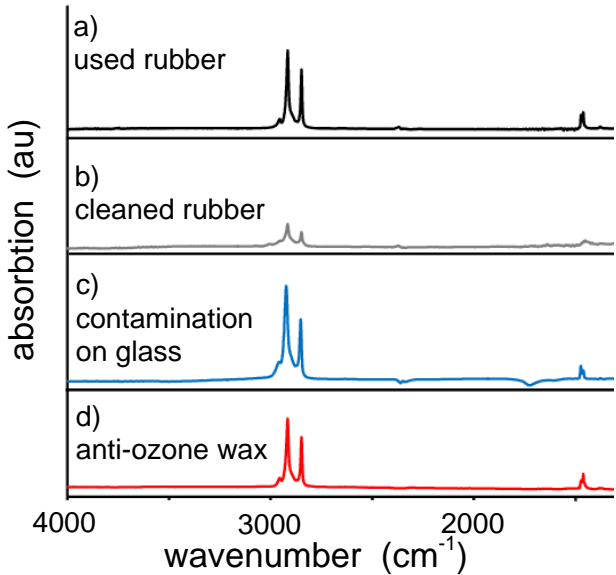}
\caption{\label{WAXozone.ps}
Surface contamination of the rubber compound A.
The infrared absorption spectra from (a) a not-cleaned rubber surface,
(b) a rubber surface cleaned with acetone, (c) the contamination film
left on a glass surface after sliding of compound A and (d) anti-ozone wax. 
}
\end{figure}

\vspace{0.3cm}
{\bf 7 Infrared spectroscopy of contamination film}

We have performed Fourier-transform infrared (FTIR) spectroscopy from the contamination film 
left on the glass surface after (sliding) contact with the rubber. In Fig. 
\ref{WAXozone.ps} we show the FTIR spectra from (a) a not-cleaned rubber surface (compound A),
and in (b) from a rubber surface cleaned with acetone, and in (c) from the contamination film
left on a glass surface after sliding of compound A. The FTIR spectra of the anti-ozone wax,
which is added to the rubber compounds, is shown in Fig. \ref{WAXozone.ps}(d). 
We conclude that the  contamination film on the glass surface (and on the rubber surface) is 
a wax, which is added to protect the rubber against ozone. The wax molecules
are mobile in the rubber matrix and even if removed from the surface of the rubber using,
e.g., acetone or hot water, it will rapidly diffuse to the surface of the rubber where it form a thin 
film which protect the rubber against the influence of ozone\cite{WAX}.

\vspace{0.3cm}
{\bf 8 Sliding friction on smooth and sandblasted hydrophilic and hydrophobic glass}

In this section we present results for the friction force when rubber blocks are sliding on
smooth (Sec. 8.1 and 8.3) and sandblasted (Sec. 8.2) glass surfaces which are either clean and
hydrophilic, or made hydrophobic
as described in Sec. 5. We also study 
how the transfer of rubber to a concrete surface may influence the friction for another 
rubber compound on the surface contaminated by the first compound (Sec. 8.4).
We note that rubber friction measurements on smooth substrates are very sensitive to 
the thickness and nature of contamination films,
and to (unavoidable) variations in the roughness of the rubber block. Thus, we observe relative large
variations in the friction force when repeating the experiments under nominally identical conditions. 
In what follows we therefore focus mainly on general trends rather than exact numerical values of the
friction coefficients. Sliding friction measurements on rough surfaces, e.g., concrete surfaces,
are much easier to perform, and result in more reproducible results.

\begin{figure}[tbp]
\includegraphics[width=0.9\columnwidth]{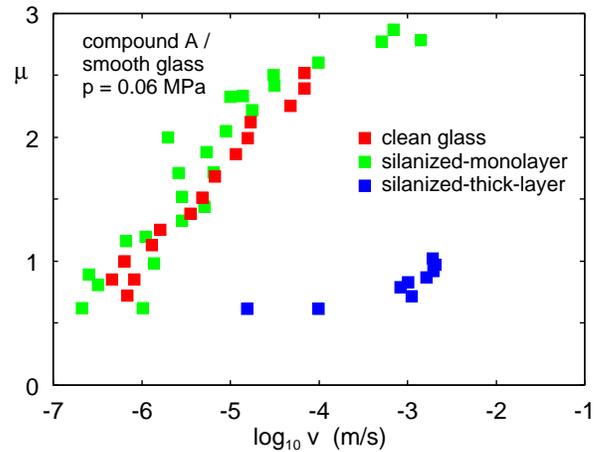}
\caption{\label{1logv.2mu.A.smooth.glass.clean.and.silanized.eps}
The measured friction coefficient for compound A on smooth silica glass plates as a function of the
logarithm of the sliding speed. 
The red symbols are the measured friction coefficient on a glass surface ultrasonic cleaned in distilled water, 
and then with distilled water. The green symbols was obtained on the same type of glass plate cleaned in the same
way and then covered by an inert monolayer film using Rain X.
The blue symbols is for a thick film of Rain X.
The nominal contact pressure $p=0.06 \ {\rm MPa}$ and the temperature $T=20^\circ {\rm C}$.
}
\end{figure}

\begin{figure}[tbp]
\includegraphics[width=0.45\textwidth]{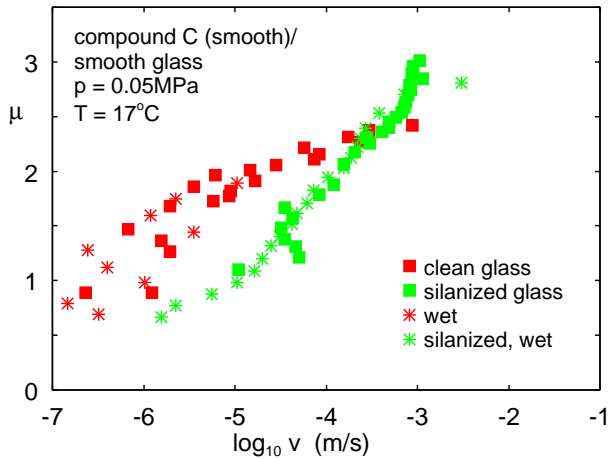}
\caption{\label{1logv.2mu.C.smooth.dry.water.eps}
The measured friction coefficient for compound C on smooth silica glass plates as a function of the
logarithm of the sliding speed. The rubber surface was cleaned in hot water and then 
dried just before start of the experiment.
The squares are for dry surfaces and the stars in water.
The red symbols are the measured friction coefficient on a glass surface cleaned by ultrasound in water.
The green symbols was obtained on the same type of glass plate cleaned in the same
way and then covered by an inert thin film using Rain X.
The nominal contact pressure $p=0.05 \ {\rm MPa}$ and the temperature $T=17^\circ {\rm C}$.
}
\end{figure}

\vspace{0.2cm}
{\bf 8.1 Sliding friction on smooth glass surfaces}

We have studied the rubber sliding friction on smooth and sandblasted
glass surfaces using the Leonardo da Vinci set up (see \cite{Tol}). 
We have used clean glass and silanized glass surfaces and studied both dry and wet
friction. 

Fig. \ref{1logv.2mu.A.smooth.glass.clean.and.silanized.eps}
shows the measured friction coefficient for the compound A on smooth silica glass plates as a function of the
logarithm of the sliding speed. 
The red symbols are the measured friction coefficient on a glass surface ultrasonic cleaned in water. 
The green symbols was obtained on the same type of glass plate cleaned in the same
way and then covered by an inert monolayer film using Rain X.
The blue symbols is for a thick film of Rain X.
The nominal contact pressure $p=0.06 \ {\rm MPa}$ and the temperature $T=20^\circ {\rm C}$.

Note that the friction coefficient before run-away is the same on the ``clean'' and
(monolayer) silanized glass surface, but extend to higher sliding speeds for the silanized surface.
Optical (reflected light at grazing incidence) pictures of the glass surfaces after the friction studies
shows that a very thin contamination film
is transferred from the rubber to the glass surface, probably already at the front
edge of the rubber-glass asperity contact regions. We propose that the friction is due to sliding on, 
or shearing of, this transfer film which is effectively pinned at the glass surface, even for the silanized glass surfaces. 

That a thin molecular film is transferred from the rubber to the countersurface even for
silanized glass was also proved by letting a small water droplet slide on the tilted glass surface: the droplet
moved quickly on the silanized glass surface in the regions which had not been in contact with the rubber, but stopped to move,
or moved in a slow and irregular way, on the regions of the glass surface which had been in sliding contact with the rubber.
As shown in Sec. 7 
the contamination film is a wax.

For the silanized glass surface which was not washed (cleaned) with distilled water after being exposed 
to the Rain X liquid, we observed
much lower friction than for the monolayer silanized surface (see Fig. \ref{1logv.2mu.A.smooth.glass.clean.and.silanized.eps}). 
We attribute this to a relative thick (maybe micrometer)
film of polydimethylsiloxane, which act as a viscous lubrication film resulting in a hydrodynamical
(but shear thinning) frictional shear stress as indicated in Fig. \ref{DiffCases.eps}(b). 

Fig. \ref{1logv.2mu.C.smooth.dry.water.eps} shows similar results as 
in Fig. \ref{1logv.2mu.A.smooth.glass.clean.and.silanized.eps} but now for the compound C.
The squares are for dry surfaces and the stars in water.
The red symbols are the measured friction coefficient on a glass surface cleaned by ultrasound in water.
The green symbols was obtained on the same type of glass plate cleaned in the same
way and then covered by an inert thin film using Rain X.

For the compound C the friction coefficient on the 
dry silanized glass surface is lower at low sliding speed than for the clean glass surface,
but larger at the highest sliding speeds.
Again we believe a very thin contamination film
(maybe just a few monolayers) is transferred from the rubber to the glass surface already at the front
edge of the rubber-glass asperity contact regions, and the friction is due to shearing of this
film. The lower friction on the silanized surface indicate that some slip may occur between the
contamination film and the silanized glass surface, at least at low sliding speed. 

For the compound C (smooth) we observe before run-away
the same friction coefficient in water as on the dry surface for both the 
clean glass surface and the silanized glass surface (see Fig. \ref{1logv.2mu.C.smooth.dry.water.eps}). 
This indicate that in both cases the water is removed from the
rubber-glass interface. This is expected for the silanized glass surface which is hydrophobic. 
In fact, Fig. \ref{AdhesionCamoothrough.eps}(a) shows that for the silanized glass
the rubber-glass adhesion in water is very strong.
However, for the clean glass
surface the adhesion in water is very weak, but the friction for low sliding speeds 
is the same as for the dry surface.
We believe this is due to a combination of fluid squeeze-out by the applied normal pressure,
and dewetting driven by the (small) adhesion observed also in water (see Fig. \ref{AdhesionCamoothrough.eps}(a)). 
Note, however, that for the clean glass surface 
the sliding in water is stable only for sliding speeds $v < 10^{-5} \ {\rm m/s}$ while in the dry state stable sliding occur up to
$v \approx 10^{-3} \ {\rm m/s}$. We interpret this difference as due to forced wetting, which is expected for 
high enough sliding speed\cite{degenneC}, and which will occur at lower sliding speed when the adhesion
in water is reduced. In forced wetting the fluid (here water)
is dragged in to the contact by the sliding motion,
resulting in viscous hydroplaning
and a decreasing friction force with increasing sliding speed and run-away. 
Forced wetting is due to a competition between a liquid invasion induced by shear
and spontaneous dewetting of the liquid driven by the (negative) spreading pressure. A similar effect (forced wetting)
was found in an earlier study with rubber sliding on a concrete surface in glycerol\cite{BoBook}.

We have repeated the experiment for compound  C (not shown)
and found more noisy results than in Fig. \ref{1logv.2mu.C.smooth.dry.water.eps}, but qualitatively the same behavior except
that for hydrophobic glass in water the run-away occurred at much lower friction coefficient than found above
($\mu \approx 1.5$ instead of $\mu \approx 2.8$ in Fig. \ref{1logv.2mu.C.smooth.dry.water.eps}). The two experiments was performed
at different environmental temperatures ($T=22^\circ {\rm C}$ and $T=17^\circ {\rm C}$),
but this is unlikely to be the origin of the different results in water. A more likely explanation is that the 
mass load (here a lead block) in the two experiments was located at slightly different position on the wood plate, 
which could tilt the plate such that the rubber-glass contact pressure 
was higher at the front edge than at the exit (in the sliding direction)
in the set-up used in the experiment reported on in Fig. \ref{1logv.2mu.C.smooth.dry.water.eps},
which could reduce or remove the liquid invasion. This tilting-problem would be absent if instead of rectangular 
rubber blocks one would use (half) cylinder shaped rubber samples.

\begin{figure}[!ht]
	\includegraphics[width=0.45\textwidth]{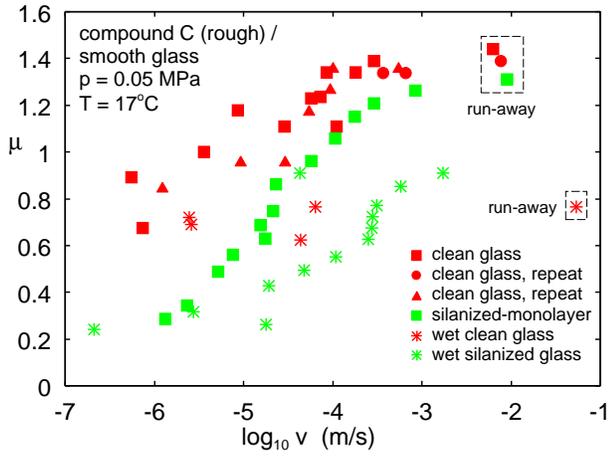}
	\caption{\label{1logv.2mu.C.rough.dry.water.1.eps}
		The measured friction coefficient for compound C on smooth silica glass plates as a function of the
		logarithm of the sliding speed. The rubber surface was roughened by sand paper.
		The squares are for dry surfaces and the stars in water.
		The red symbols are the measured friction coefficient on a glass surface cleaned by ultrasound in water.
		The green symbols was obtained on the same type of glass plate cleaned in the same
		way and then covered by an inert thin film using Rain X.
		The nominal contact pressure $p=0.05 \ {\rm MPa}$ and the temperature $T=17^\circ {\rm C}$.}
\end{figure}

Fig. \ref{1logv.2mu.C.rough.dry.water.1.eps}
shows the measured friction coefficient for compound C (rough) on smooth silica glass plates as a function of the
logarithm of the sliding speed. The rubber surface was roughened by sand paper.
The squares and triangles are for dry surfaces and the stars in water.
The red symbols are the measured friction coefficient on a glass surface cleaned by ultrasound in water.
The green symbols was obtained on the same type of glass plate cleaned in the same
way and then covered by an inert thin film using Rain X.

\begin{figure}[!ht]
\includegraphics[width=0.8\columnwidth]{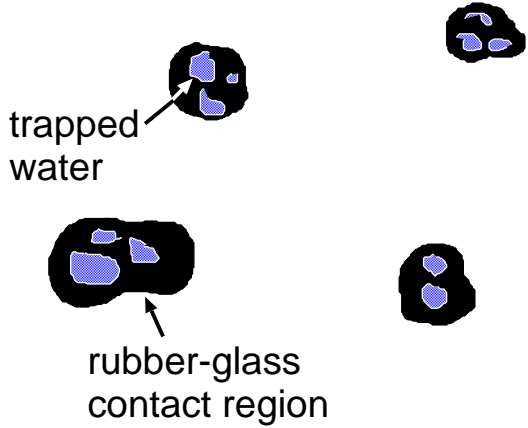}
\caption{\label{TrappedWater.eps} 
		Rubber-glass asperity contact regions in water. At short enough length scale
		the contact pressure becomes so high that the contact area (locally) percolate.
		When the contact is formed in water, pressurized water may occur in the sealed-off
		regions. In this case part of the external load will be carried by the water resulting in
		a reduced rubber-glass contact area and a reduced sliding friction.}
\end{figure}

Note that for the dry clean and silanized glass surfaces the velocity dependency of the friction coefficient 
for the roughened rubber surface (Fig. \ref{1logv.2mu.C.rough.dry.water.1.eps}) is similar to for the
smooth rubber surface (Fig. \ref{1logv.2mu.C.smooth.dry.water.eps}),
but the friction coefficient is smaller for the rough rubber surface as expected because of the smaller area of real
contact. However, in water the results are very different. Thus for the clean glass surface 
the friction is only weakly velocity dependent, and similar in magnitude to the sliding friction on the 
dry surface at the lowest sliding speed. For the hydrophobic glass surface the friction in water 
is smaller than for the dry state but exhibit a similar velocity dependency. The reduction in the
friction coefficient in water for the rough rubber surface may be due to water trapped in 
sealed-off regions which carry part of the external load, and hence reduces the area of real contact.
In the present case the applied squeezing pressure is not high, but the local pressure in asperity contact
regions will be high and, at short enough length scale, the
contact area may percolate in the asperity contact regions, 
resulting in water filled sealed-off regions (see Fig. \ref{TrappedWater.eps}).


\vspace{0.2cm}
{\bf  8.2 Sliding friction on sandblasted glass surfaces}

\begin{figure}[!hb]
	\includegraphics[width=0.9\columnwidth]{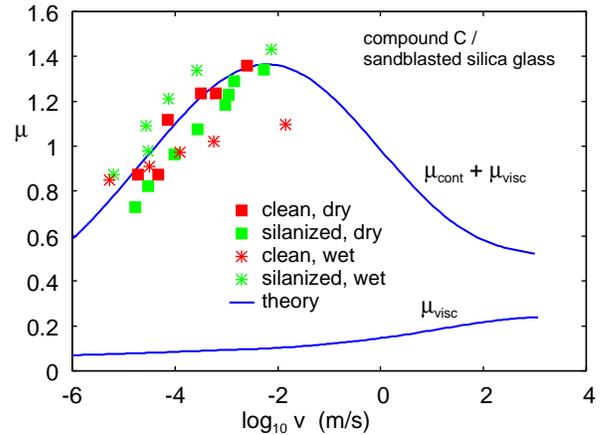}
	\caption{\label{1logv.2mu.sandblastedglass.C.with.thory.eps}
		The measured friction coefficient for 
		compound C on sandblasted silica glass plates as a function of the
		logarithm of the sliding speed. The squares are for dry surfaces and the stars in water.
		The red symbols are the measured friction coefficient on a glass surface ultrasonic cleaned in water. 
		The green symbols was obtained on the same type of glass plate cleaned in the same
		way), and then covered by an inert thin film using Rain X.
		The nominal contact pressure $p=0.05 \ {\rm MPa}$ and the temperature $T=22^\circ {\rm C}$.
	}
\end{figure}

Fig. \ref{1logv.2mu.sandblastedglass.C.with.thory.eps}
shows the measured friction coefficient for 
compound C sliding on a sandblasted silica glass plate as a function of the
logarithm of the sliding speed. The squares are for dry surfaces and the stars in water.
The red symbols are the measured friction coefficient on a glass surface ultrasonic cleaned in water 
and then with distilled water. The green symbols was obtained on the same type of glass plate cleaned in the same
way, and then covered by an inert monolayer film using Rain X. 

\begin{figure}[!hb]
	\includegraphics[width=0.9\columnwidth]{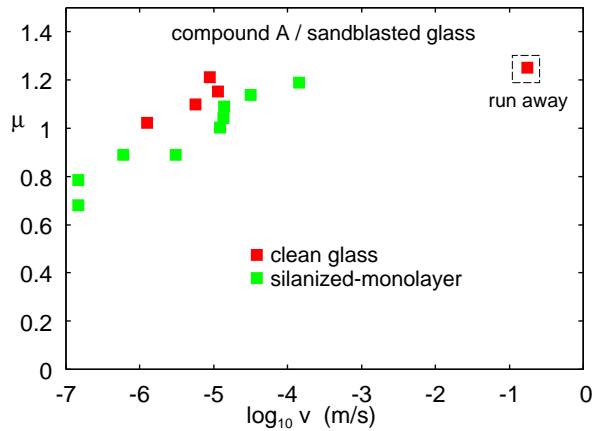}
	\caption{\label{1logv.2mu.A.sandblasted.eps}
		The measured friction coefficient for 
		compound A on sandblasted silica glass plates as a function of the
		logarithm of the sliding speed. 
		The red squares are the measured friction coefficient on a glass surface ultrasonic cleaned in water. 
		The green squares was obtained on the same type of glass plate cleaned in the same
		way), and then covered by an inert thin film using Rain X.
		The nominal contact pressure $p=0.11 \ {\rm MPa}$ and the temperature $T=16^\circ {\rm C}$.
	}
\end{figure}

The friction coefficient is very similar for the clean and silanized
glass surfaces in spite of the fact that the work
of adhesion for the (smooth) clean glass is twice as high as for the silanized glass
surface (see Fig. \ref{AdhesionCamoothrough.eps}). This indicate again the transfer of a 
contamination film to the glass surface from the rubber
and that the film is effectively pinned at the glass interface even for the  
silanized glass surface. For sliding in water, for the low sliding speed prevailing in the experiments, 
the water is likely to be squeezed out from the asperity contact regions. 
In this case it appears as if no islands of trapped (sealed-off) water 
occur. This may be due to the increased surface roughness which result 
in a smaller contact area and in larger non-contact channels where the water can be removed. 

Fig. \ref{1logv.2mu.A.sandblasted.eps}
shows the measured friction coefficient for 
compound A on sandblasted silica glass plates as a function of the
logarithm of the sliding speed. 
Again the rubber friction on the clean glass surface and the silanized glass is nearly the same.
Finally we note that on the sandblasted glass surface there will be a viscoelastic contribution to
the friction, which does not exist on the smooth glass surfaces even when the rubber surface is rough.



\begin{figure}[tbp]
\includegraphics[width=0.47\textwidth,angle=0]{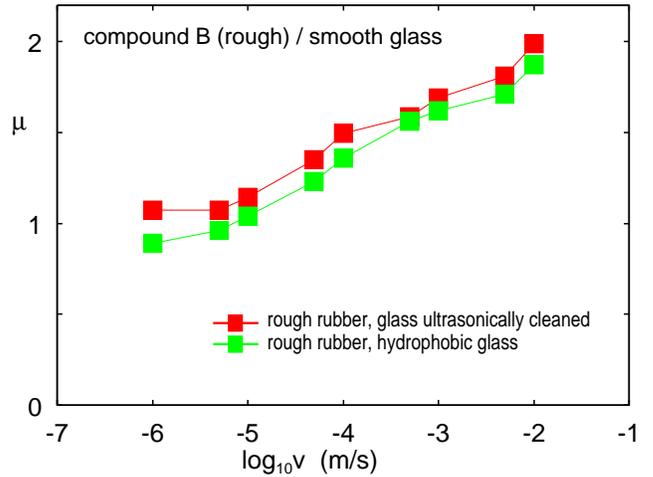}
\caption{
The friction coefficient as a function of the logarithm of the sliding speed
for the compound B. 
The red squares are on a dry glass surface cleaned by acetone and 
isopropanol and then ultrasonic cleaned in distilled water. The green squares are for a hydrophobic
(silanized) glass. The rubber block surface is roughened by sand paper. 
The nominal contact pressure $p=0.06 \ {\rm MPa}$ and the temperature $T=20^\circ {\rm C}$.
}
\label{ROughAll.eps}
\end{figure}

\begin{figure}[tbp]
\includegraphics[width=0.47\textwidth,angle=0]{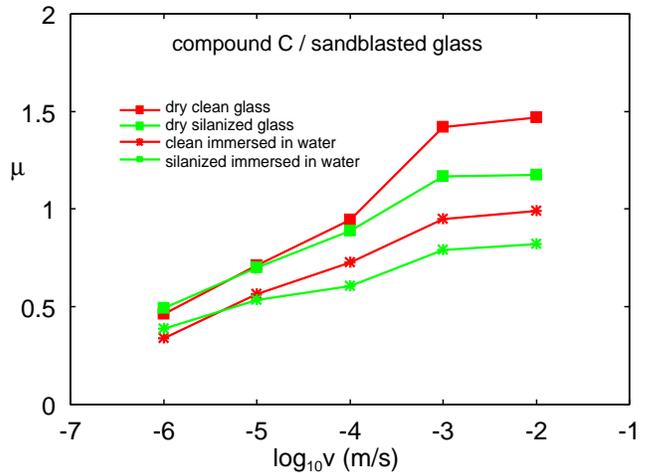}
\caption{
The friction coefficient as a function of the logarithm of the sliding speed
for the compound C on sandblasted glass. 
The red squares are on a dry glass surface cleaned by acetone and 
isopropanol and then ultrasonic cleaned in distilled water. The green squares are for a hydrophobic
(silanized) glass. The square symbols are for dry contact and the stars in water.
The nominal contact pressure $p=0.06 \ {\rm MPa}$ and the temperature $T=20^\circ {\rm C}$.
}
\label{compoundLTFreezerhydrophilicvshydrophobic.1.eps}
\end{figure}

\vspace{0.1cm}
{\bf 8.3 Results obtained using a linear friction slider}

We have also performed friction studies using a (low-temperature) linear 
friction tester (see Ref. \cite{Tol}). 
Fig. \ref{ROughAll.eps} shows the friction coefficient as a function of the logarithm of the sliding speed
for the compound B 
on a dry smooth glass surface cleaned by acetone and 
isopropanol and then ultrasonic cleaned in distilled water (red), and on a hydrophobic
(silanized) glass surface (green). The rubber block surface is roughened by sand paper, 
which removes a thin surface layer of rubber, which may have different
properties from the bulk rubber.

Note that in the present case the dry ``clean'' glass surface
exhibit nearly the same friction as for the silanized surface,
as also observed in the Leonardo da Vinci study (not shown). 

Fig. \ref{compoundLTFreezerhydrophilicvshydrophobic.1.eps} 
shows the friction coefficient as a function of the logarithm of the sliding speed
for the compound C on sandblasted glass. 
The red squares are on a dry glass surface cleaned by acetone and 
isopropanol and then ultrasonic cleaned in distilled water. The green squares are for a hydrophobic
(silanized) glass. The square symbols are for dry contact and the stars in water.
The friction coefficients are similar to obtained in Fig. \ref{1logv.2mu.sandblastedglass.C.with.thory.eps} 
using the Leonardo da Vinci set up
but at the higher sliding speeds the hydrophobic glass gives slightly smaller friction than the clean glass,
both in the dry state and in water. Also, in the water the friction is smaller than in the dry state.
We believe the difference between the results in water in Fig. \ref{1logv.2mu.sandblastedglass.C.with.thory.eps} and 
Fig. \ref{compoundLTFreezerhydrophilicvshydrophobic.1.eps} is due to differences in the stiffness of the two experimental set-up
and a (related) slight tilting of the rubber blocks which effect the removal of the water between the two surfaces.

\begin{figure}[tbp]
\includegraphics[width=0.47\textwidth,angle=0]{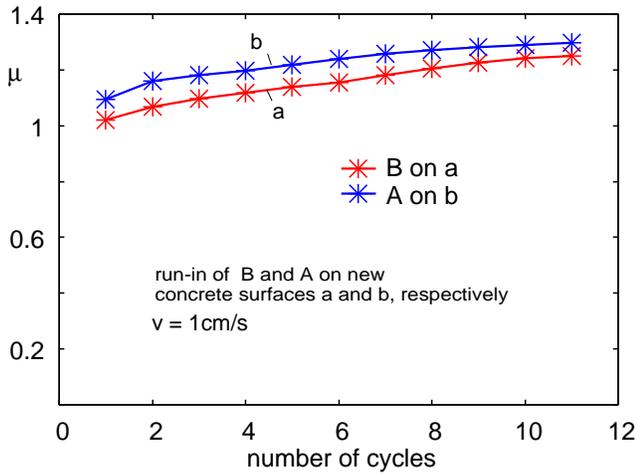}
\caption{
The friction coefficient at the sliding speed $v=1 \ {\rm cm/s}$ as a function of the number
of sliding cycles for compound B (red) and A (blue) sliding on two different concrete
blocks denoted {\bf a} and {\bf b}, respectively. 
}
\label{RuninBandA.eps}
\end{figure}

\begin{figure}[!ht]
\includegraphics[width=0.9\columnwidth]{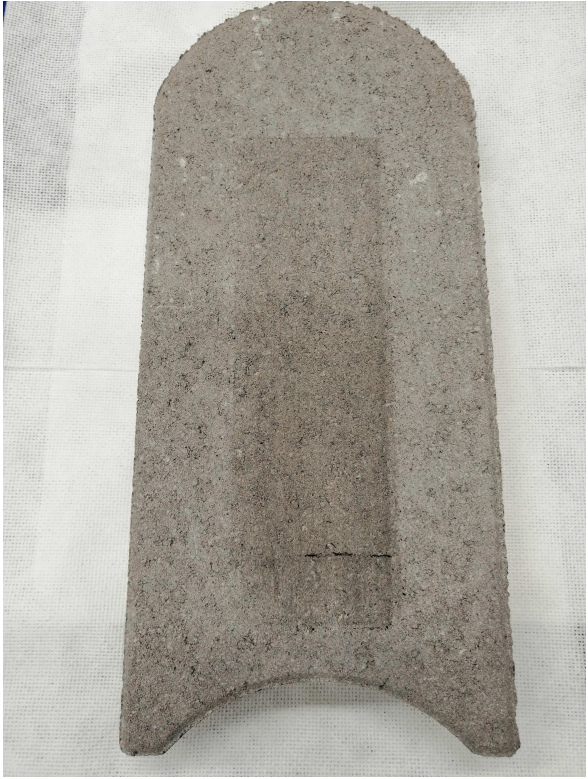}
\caption{\label{stone.eps}
Picture of concrete block {\bf a} after run-in of compound B.
}
\end{figure}

\begin{figure}[tbp]
\includegraphics[width=0.47\textwidth,angle=0]{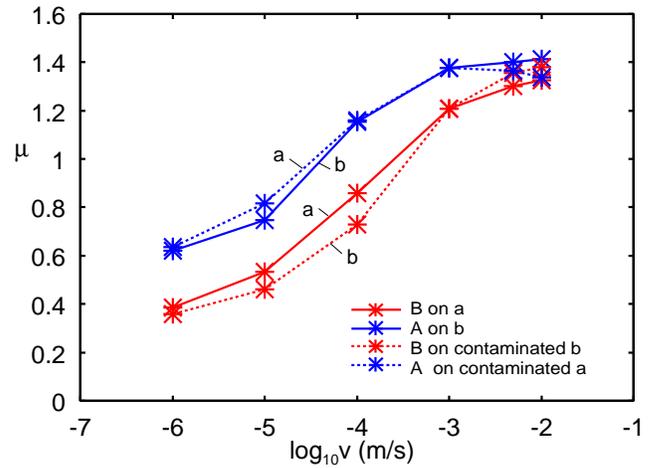}
\caption{
The friction coefficient as a function of the logarithm of the sliding speed.
The solid lines are for the compound B (red line) and A (blue line) sliding
on two different concrete blocks ({\bf a} and {\bf b}, respectively) 
after run-in on the same concrete blocks (resulting in ``contaminated'' concrete surfaces). 
The red dashed line is for compound B sliding on the concrete block {\bf b} after
it was contaminated by the compound A. The blue dashed line is for compound A 
sliding on the concrete block {\bf a} after it was contaminated by the compound B.
}
\label{Contaminationsliding.eps}
\end{figure}

\vspace{0.3cm}
{\bf 8.4 Sliding friction on contaminated concrete}

It is often stated in F1-racing that transfer of rubber to the racing track can have a strong influence on the sliding friction
or grip. In particular, when changing to new tires with a different rubber tread compound the friction can be reduced compared
to clean road surface. In an earlier study we observed a strong drop in the friction
when a rubber block was slid on a concrete surface contaminated by first sliding a 
block made from another rubber compound on the concrete surface\cite{Tol}. 

We have performed a set of experiments to study the role of rubber contamination
films on rubber friction on a concrete surface. In the experiments we first run-in 
the rubber compounds B and A on two different concrete surfaces {\bf a} and {\bf b}.
The run-in consisted of 11 forwards and backwards sliding events (sliding speed
$v=1 \ {\rm cm/s}$) of compound B on surface {\bf a}, and similar for compound A on surface {\bf b}. 
Fig. \ref{RuninBandA.eps} shows the friction coefficient as a function of the number
of sliding cycles for compound B (red) and A (blue) sliding on the two different concrete
surfaces {\bf a} and {\bf b}. The two concrete surfaces are nominally identical so the small
difference in the friction observed may be attributed to the different rubber compounds.
Note that during run-in there is an increase in the friction by $\sim 20\%$ in both cases.
This may be due to removing a thin skin layer on the rubber blocks or due to the influence
on the friction by a thin contamination film deposited on the concrete surface.
Fig. \ref{stone.eps} shows a picture of the concrete block {\bf a} after run-in of compound B.

Next we measured the friction force for compound B on surface {\bf a} for sliding speeds
from $1 \ {\rm \mu m/s}$ to $1 \ {\rm cm/s}$, and the same experiment was carried out for compound A on the surface {\bf b}.
The results are given by the solid red and blue lines in Fig. \ref{Contaminationsliding.eps}. Next we measured
the friction for the same velocity interval for compound B on the concrete surface {\bf b} 
contaminated by compound A (red dashed line in Fig. \ref{Contaminationsliding.eps}). Finally, we performed
a similar experiment for compound A on the concrete surface {\bf a} already contaminated by the
compound B (blue dashed line in Fig. \ref{Contaminationsliding.eps}). As shown in the figure, within the
noise of the measured data, the friction coefficient is nearly the same for compound B
when sliding on surface {\bf a} contaminated by compound B, as on surface {\bf b} contaminated
by the compound A (red lines). The same is true for compound A (blue lines).
However, in all cases the compound A gives higher friction than the compound B, but the difference
is rather small at the highest sliding speed, where the difference is consistent with the friction observed
during run-in.

We have shown that the friction changes very little when sliding compound B on the concrete surface
first contaminated by compound A, and the same was observed also in the opposite case. This indicate
that either the contamination film is sheared as in Fig. \ref{DiffCases.eps}(b), in which case the frictional shear stress
may be independent of the rubber compound used, or the chain interdiffusion mentioned above may occur 
in a similar way for the compound B in contact
with the transfer film from compound A as with the transfer film from compound B. Alternatively,
during sliding of compound B on the concrete surface contaminated by the compound A there is a new transfer film 
of compound B deposited on the concrete surface which could result in a contact which is effectively the same in
both cases.


\vspace{0.3cm}
{\bf 9 Discussion}

We have found above that for smooth and sandblasted glass surfaces the friction force for compounds A and B
is nearly the same on the ``clean'' glass surfaces as on the silanized glass surfaces. 
For compound C the same is true on the sandblasted glass surface in the Leonardo da Vinci set up, 
but for the smooth glass surface the silanized surface exhibit lower friction at low sliding speed.
Optical pictures show that molecules are transferred from the rubber to the glass surface,
and FTIR spectra shows that the contamination film is a wax.
If we assume that negligible slip occur at the interface between the contamination film and
the clean or silanized glass surface, then the friction force will be the same on both glass surfaces.
Only if we assume this to be the case is it possible to explain the great similarity
in the observed friction coefficient on the two different surfaces for compounds A and B. 
Thus, for these compounds in both cases we propose that
most of the slip may occur at the interface between the rubber and the contamination
film [see Fig. \ref{DiffCases.eps}(a)], or result from shearing the lubrication film as in
Fig. \ref{DiffCases.eps}(b). On the silanized surface the contamination film binds weaker than on the clean surface but,
nevertheless, it appear that in many cases there is negligible slip between the contamination film and the silanized glass surface.
This differ from the result of a recent study where slip was observed between a rubber stopper and a glass barrel
with baked-on silicone oil\cite{Paolo}. 

In wet conditions the silanazed glass surface result in a higher 
maximum friction coefficient than the hydrophilic glass surface.
We interpret this as resulting from forced wetting. In forced wetting the water
is dragged into the contact by the sliding motion,
resulting in viscous hydroplaning
and a decreasing friction force with increasing sliding speed. 
Forced wetting is due to a competition between a liquid invasion induced by shear
and spontaneous dewetting of the liquid driven by the (negative) spreading pressure. 

\begin{figure}
\includegraphics[width=0.50\textwidth,angle=0.0]{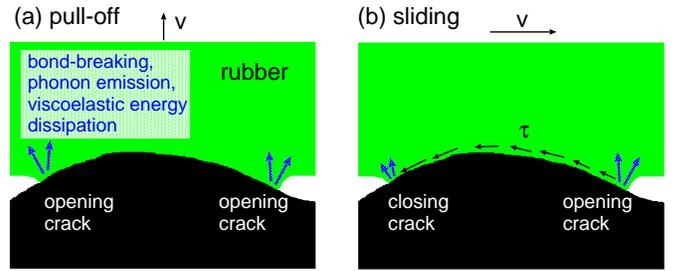}
\caption{\label{sheststress.eps}
There is in general no simple relation between adhesion (pull-off force) and sliding friction. (a) The pull-off force
depends on breaking the bonds in the normal direction at the edges of the contact region (opening crack
propagation). (b) The friction force depends on energy dissipation at the opening and closing crack tips, and in addition
on shearing the area of real contact, i.e., on processes occurring
everywhere within the contact region. In many cases this latter contribution will give the most important
contribution to the friction force\cite{Gert}.
}
\end{figure}

For the clean glass surface the drop in the work of adhesion between dry contact and contact 
is in water (see Fig. \ref{1number.2w.APOLLO.eps})
is much higher than the change in the
the friction force between dry and wet condition. Hence,
there is no simple relation between adhesion (pull-off force) and sliding friction. 
This is easy to understand (see Fig. \ref{sheststress.eps}) because the pull-off force
depends on breaking the bonds in the normal direction at the edges of the contact region, while the
frictional shear stress depends mainly on shearing the contact area, i.e., on processes 
occurring everywhere within the contact region. 
The latter may not even ``know'' that there is fluid (water) outside the asperity contact regions,
which may strongly reduce the bond-breaking at the edges of the contact regions.

When the substrate has surface roughness (as in Fig. \ref{sheststress.eps}) there will also be a contribution
to the friction from the opening crack tip on the exit side of the sliding rubber-substrate contact regions\cite{PNAS}, but
for small contacts the latter contribution will be reduced due to finite 
size effects\cite{EPLsmall}, and experiments\cite{Gert} indicate that, in most practical applications,
the main contribution to the friction arises from the inner region of the asperity contact regions.

\vspace{0.3cm}
{\bf 10 Summary and conclusion}

In this article we have presented results for rubber adhesion and friction against glass in dry and wet conditions.
The surface energy and topography of the glass was changed by application of hydrophobic coating (silanization) 
and sandblasting.

We found in many cases that changing the surface energy of the glass by silanizing does not 
affect the magnitude of rubber friction on smooth and sandblasted glass. 
We believe that this is due to 
sliding on, or shearing of, a contamination film derived from the rubber, with negligible slip at 
the glass-contamination film interface, even for the silanized glass surfaces. This indicate that for dry surfaces
the surface energy of the substrate may be irrelevant for rubber friction in many practical cases. 
Infrared light absorption measurements shows that the contamination film is 
a wax, which was added to the rubber compounds to protect against ozone.

The magnitude of rubber friction for smooth glass depends on whether the rubber is smooth or roughened prior
to friction measurements. Smooth rubber showed higher friction as compared to the roughened rubber. 
We attribute the decrease in the friction for roughened rubber to the reduced area of contact.

When rubber slides on a clean smooth glass in the wet state we observe runaway instabilities at low sliding speed.
We attribute this to forced wetting, where water is dragged into the sliding interface during sliding motion.

When the rubber or countersurface are rough the rubber friction is lower in water than that in the dry state.
We attribute this to a trapped islands of pressurized water,
which carry part of the external load and hence reduces the area of real contact.
The trapped islands of water form at short enough length scale 
(observed at high enough magnification),
where the rubber-substrate contact area percolate.

We have shown that the transfer of rubber to a concrete track 
has only a small influence the friction for another 
rubber compound and vice verse. However, more studies of this, using rubber compounds with bigger difference
in the chemical composition, are necessary in order to determine how general this conclusion is.

Our study demonstrate that there is no simple relation between adhesion and friction.
Thus, adhesion is due to (vertical) detachment processes
at the edge of the contact regions (opening crack propagation), 
while friction in many cases is determined 
mainly by (tangential) stick-slip instabilities within the whole
sliding contact.

\end{document}